\documentclass[usenatbib]{emulateapj}
\usepackage{graphicx}
\usepackage[flushleft]{threeparttable}
\usepackage[usenames,dvipsnames,svgnames,table]{xcolor}
\usepackage{hyperref}
\definecolor{darkblue}{rgb}{0.0,0.0,0.9}
\hypersetup{colorlinks,breaklinks,
            linkcolor=darkblue,urlcolor=darkblue,
            anchorcolor=darkblue,citecolor=darkblue}
\usepackage{amsmath,amssymb}
\usepackage{amsmath}

\usepackage{natbib}
\begin{document}

\def\etal{et al.\ \rm}
\def\ba{\begin{eqnarray}}
\def\ea{\end{eqnarray}}
\def\etal{et al.\ \rm}
\def\Fdw{F_{\rm dw}}
\def\Tex{T_{\rm ex}}
\def\Fdis{F_{\rm dw,dis}}
\def\Fnu{F_\nu}
\def\WD{\rm WD}

\newcommand{\ra}{r_\mathrm{a}}
\newcommand{\rb}{r_\mathrm{b}}
\newcommand{\rc}{r_\mathrm{c}}
\newcommand{\md}{\mathrm{d}}
\newcommand{\me}{\mathrm{e}}
\newcommand{\mi}{\mathrm{i}}
\newcommand{\rp}{r_\mathrm{p}}
\newcommand{\fmer}{f_\mathrm{m}}
\newcommand{\rt}{\mathrm{t}}
\newcommand{\fbara}{\overline{f}_0^\mathrm{a}}
\newcommand{\fbarb}{\overline{f}_0^\mathrm{b}}
\newcommand{\fdotnow}{(\md f_\mathrm{m}/\md t)|_{12\,\mathrm{Gyr}}}
\newcommand{\fnow}{f_\mathrm{m}(12\,\mathrm{Gyr})}
\def\p{\partial}
\newcommand{\Rg}{\mathbf{R}_\mathrm{g}}
\newcommand{\epsGR}{\epsilon_\mathrm{GR}}
\newcommand{\nn}{\nonumber}

\newcommand\cmtrr[1]{{\color{red}[RR: #1]}}
\newcommand\cmtch[1]{{\color{red}[CH: #1]}}


\title{Compact object binary mergers driven by cluster tides: a new channel for LIGO/Virgo gravitational wave events}

\author{Chris Hamilton\altaffilmark{1,$\dagger$} \& Roman R. Rafikov\altaffilmark{1,2}}
\altaffiltext{1}{Department of Applied Mathematics and Theoretical Physics, University of Cambridge, Wilberforce Road, Cambridge CB3 0WA, UK}
\altaffiltext{2}{Institute for Advanced Study, Einstein Drive, Princeton, NJ 08540}
\altaffiltext{$\dagger$}{ch783@cam.ac.uk}


\begin{abstract}
The detections of gravitational waves produced in mergers of binary black holes (BH) and neutron stars (NS) by LIGO/Virgo have stimulated interest in the origin of the progenitor binaries. Dense stellar systems --- globular and nuclear star clusters --- are natural sites of compact object binary formation and evolution towards merger. Here we explore a new channel for the production of binary mergers in clusters, in which the tidal field of the cluster secularly drives the binary to high eccentricity (even in the absence of a central massive black hole) until gravitational wave emission becomes important. We employ the recently developed secular theory of cluster tide-driven binary evolution to compute present day merger rates for BH-BH, NS-BH and NS-NS binaries, varying cluster potential and central concentration of the binary population (but ignoring cluster evolution and stellar flybys for now). Unlike other mechanisms, this new dynamical channel can produce a significant number of mergers out to cluster-centric distances of several pc. For NS-NS binaries we find merger rates in the range $0.01-0.07$ Gpc$^{-3}$ yr$^{-1}$ from globular clusters and $0.1-0.2$ Gpc$^{-3}$ yr$^{-1}$ from cusped nuclear clusters. For NS-BH and BH-BH binaries we find small merger rates from globular clusters, but a rate of $0.1 - 0.2$ Gpc$^{-3}$ yr$^{-1}$ from cusped nuclear clusters, contributing to the observed LIGO/Virgo rate at the level of several per cent. Therefore, cluster tide-driven mergers constitute a new channel that can be further explored with current and future gravitational wave detectors.
\end{abstract}





\section{Introduction}


The detection of gravitational waves (GWs) produced in mergers of binaries composed of compact objects --- black holes (BHs) and neutron stars (NSs) --- by the LIGO/Virgo collaboration \citep{LIGO2018,Teja2019} naturally raised the question of the origin and evolutionary pathways of these systems. While individual NSs and BHs are the known end states of the collapse of massive stars, the dominant mechanisms by which they combine into small-separation binaries and finally merge are still uncertain. Small separation is the key, since e.g. a circular binary composed of two $30M_\odot$ BHs can merge via GW emission in a Hubble time only if it has an initial semi-major axis of $\lesssim 0.2$ au. 

One possible channel by which small semi-major axis can be acheived is stellar evolution of binaries composed of two massive stars,
e.g. through a common-envelope phase \citep{Paczynski1971,Tutukov,Iben1993,Taam2000,Kalogera2007,Belczynski2016}, or through chemically homogeneous evolution as a result of rapid rotational mixing \citep{Mandel2016}. A different channel is provided by secular dynamics of compact object binaries in triples \citep{Antonini2014,Antonini2016b,Silsbee2017,Liu2017}: an inner binary can be torqued by its tertiary companion into performing Lidov-Kozai (LK) oscillations \citep{Lidov1962,Kozai1962}, forcing it to very high eccentricity and thereby boosting the rate of gravitational wave (GW) emission and shrinking its semi-major axis. 

Dense stellar clusters provide several alternative avenues for the formation of compact object binaries. Three- and four-body encounters in the dense environments of clusters greatly enhance the binary NS formation rate \textit{dynamically}: the abundance per unit mass of low-mass X-ray binaries is around $10^2$ times higher in globulars, and $10^3$ times higher in the central parsec of the Galaxy, than it is in the Galactic field \citep{Katz1975,Clark1975,Generozov2018}. Similarly, BH-BH binaries should form dynamically in cluster cores provided the BHs are retained in their clusters at birth  \citep{PortegiesZwart2000,OLeary2006,Rodriguez2016a,Antonini2016b}. This possibility is supported by the recent discovery of a detached binary consitisting of a BH and a main-sequence turnoff star in the globular cluster NGC 3201 \citep{Giesers2018}.  

As the majority of dynamically formed relativistic binaries are too wide to merge via GW emission within a Hubble time, it is not enough to explain how they form: one must also explain how they shrink. Frequent stellar encounters can harden binaries in cluster cores, leading to eventual mergers that might occur after the binary is ejected from the cluster  \citep{Antonini2016a,Leigh2018}. For binaries in nuclear star clusters, a central supermassive black hole (SMBH), if present, can play the role of the tertiary driving LK oscillations and orbital decay (e.g. \citealt{Antonini2012,Petrovich2017,Hamers2018_VRR}), similar to triples in the field. 

However, so far no studies have accounted for the direct effect of the tidal field of the dense cluster to which the binary belongs on the evolution of its orbital elements (although studies of Oort comet dynamics \textit{have} routinely accounted for the Galactic tide --- see e.g. \citealt{Heisler1986,Matese1996}). Recently in \citet{Hamilton2019a,Hamilton2019b} --- hereafter `Paper I' and `Paper II' respectively --- we showed that the smooth tidal potential of a host star cluster can drive wide binaries to perform LK-like secular eccentricity oscillations on timescales that could be relevant for the production of LIGO sources. In this Letter we explore the consequences of this mechanism for the merger rate of compact object binaries (\S \ref{sec_mergerrates}), under the simplifying assumption that they orbit spherical star clusters and their dynamics are driven only by the smooth, time-independent cluster potential (i.e. we neglect the effects of flyby encounters, dynamical friction, etc., which are discussed in \S \ref{sec_discussion}).



\section{Dynamical framework} 
\label{sec_setup}


We consider a compact object binary with component masses $m_1$, $m_2$ orbiting in a fixed smooth background potential $\Phi$ of a spherically symmetric star cluster (globular or nuclear). Spherical symmetry implies that the binary's `outer' barycentric orbit is confined to a plane, which we define as the $(X,Y)$ plane, and typically densely fills an axisymmetric annulus in this plane with inner and outer radii $(r_\mathrm{p}, r_\mathrm{a})$.  
The binary's `inner' orbit (i.e. the motion of $m_1$ and $m_2$ around each other) is described by the usual orbital elements: semi-major axis $a$, eccentricity $e$, inclination $i$ (measured relative to the outer orbital plane), longitude of the ascending node $\Omega$ (relative to the $X$ axis, which is fixed in the cluster frame) and argument of pericentre $\omega$. 

We showed in Paper I that the dynamical evolution of the binary's inner orbital elements is governed by the secular (`doubly-averaged', hereafter DA) perturbing Hamiltonian\footnote{The Hamiltonian is `doubly-averaged' in the sense that it is derived by integrating first over the inner Keplerian orbit of the binary components about their common barycentre, and then again over many outer orbits of the binary itself around its host cluster.}
\begin{align} 
H = \frac{Aa^2}{8}( H_1^* + H_\mathrm{GR}^*), 
\label{HMt} 
\end{align} 
where $A$ is a constant (with units of s$^{-2}$). Here $H_1^*$ and $H_\mathrm{GR}^*$ are the dimensionless Hamiltonians accounting for quadrupole-order cluster tides and general relativistic (GR) pericentre precession, respectively:
\begin{align}
    &H_1^* = (2+3e^2)(1-3\Gamma\cos^2i)-15\Gamma e^2 \sin^2 i \cos 2\omega, 
    \label{H1star} \\ 
    &H_\mathrm{GR}^* = -\epsilon_\mathrm{GR}(1-e^2)^{-1/2}, 
    \label{HGR}
\end{align}
where $\Gamma$ is a dimensionless parameter discussed below,  and the relative strength of GR precession is measured by another dimensionless parameter
\begin{align} 
 \label{eq:epsGRformula}
\epsilon_\mathrm{GR} &\equiv \frac{24G^2(m_1+m_2)^2}{c^2Aa^4} \\ \nn &=  0.258 \times \left( \frac{A^*}{0.5}\right)^{-1}\left( \frac{M}{10^5M_\odot}\right)^{-1}\left( \frac{b}{\mathrm{pc}}\right)^{3}  \\ &\times  \left( \frac{m_1+m_2}{M_\odot}\right)^{2}  \left( \frac{a}{20 \, \mathrm{au}}\right)^{-4}. \label{eq:epsGRnumerical}
\end{align} 
In the numerical estimate \eqref{eq:epsGRnumerical} we have assumed that the binary is orbiting a spherical cluster with scale radius $b$ and total mass $M$ and introduced a futher dimensionless parameter $A^* \equiv A/(GM/b^3)$ (which is a natural scaling for $A$, see Paper I). 

\begin{figure}
\centering
\includegraphics[width=0.8\linewidth]{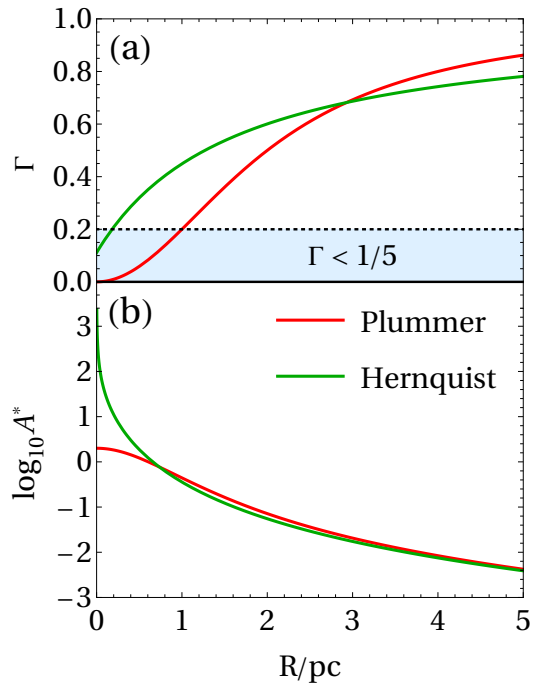}
\caption{Plots of the parameters $\Gamma$ and $A^*$ (see text) assuming the binary is on a circular outer orbit of radius $R$ in Plummer (red) and Hernquist (green) potentials each with half-mass radius $1.31\mathrm{pc}$. For initial inclinations close to $90^\circ$, high eccentricity excitation is readily achieved when $\Gamma > 1/5$, but is much rarer when $\Gamma < 1/5$ (shaded region in panel (a)).}
\label{fig:GammaOfR}
\end{figure}

The parameter $A$ (or $A^*$) measures the strength of the tidal torque and sets the timescale for secular evolution, $t_{\rm sec}\sim n/A$ (Paper II), where $n=[G(m_1+m_2)/a^3]^{1/2}$ is the binary's mean motion. The value of $A^*$ is fully determined by stipulating the cluster potential $\Phi$ and the peri/apocentre $(\rp,\ra)$ of the binary's outer orbit. In Figure \ref{fig:GammaOfR}b we plot $A^*(R)$ assuming a circular outer orbit of radius $R$ in Plummer (cored) and Hernquist (cusped, with density $\rho \propto r^{-1}$ for $r\to 0$) potentials
\begin{align} 
\Phi_\mathrm{Plum}(r) = -\frac{GM}{\sqrt{b_\mathrm{Plum}^2+r^2}}, \,\,\,\,\,\,\, \Phi_\mathrm{Hern}(r) = -\frac{GM}{b_\mathrm{Hern}+r},
\label{eq:Plummer} 
\end{align}
where $M$ is the total mass of the cluster and $b_\mathrm{Plum/Hern}$ are the corresponding scale radii. We choose $b_\mathrm{Plum} = 1\mathrm{pc}$ and $b_\mathrm{Hern} = 0.544\mathrm{pc}$ respectively so that the two potentials have the same half mass radius $r_\mathrm{h} = 1.31\mathrm{pc}$.  

The tidal Hamiltonian \eqref{H1star} differs from the dimensionless LK Hamiltonian only through the parameter $\Gamma$, which is the key characteristic of cluster tide-driven secular dynamics. Its value is also fully determined (like that of $A$) by stipulating $\Phi$ and $(\rp,\ra)$. For binaries in realistic spherical clusters we always have $0 < \Gamma \leq 1$ (Paper I), while the LK Hamiltonian is exactly recovered when $\Gamma=1$. Figure \ref{fig:GammaOfR}a shows the profiles of $\Gamma(R)$ in clusters with Plummer and Hernquist potentials.

Papers I \& II focused almost exclusively on exploring the dynamics arising from the tidal Hamiltonian \eqref{H1star}, ignoring GR precession. A key conclusion of these studies was that high eccentricities can be reached by binaries sufficiently inclined with respect to their outer orbital plane for a range of $\Gamma$ values. However, due to a bifurcation in the dynamical phase portrait, very high eccentricity is much more readily achieved by binaries with $\Gamma>1/5$ than those with $\Gamma < 1/5$. Therefore, according to Figure \ref{fig:GammaOfR}, high eccentricity should be easier to reach in cusped (e.g. Hernquist) clusters.  

To compute merger rates due to cluster tides the `doubly-averaged' calculations of Papers I \& II need to be extended by accounting for two additional effects. First, GR precession (embodied in the term \eqref{HGR}) typically acts to promote faster evolution of $\omega$, quenching the cluster tide-driven eccentricity oscillations (see e.g. \citet{Fabrycky2007} for a discussion in the LK limit, $\Gamma=1$). Reaching high $e$ in the presence of GR precession necessarily requires a sufficiently dense/massive cluster. Empirically, we find that one should not expect high eccentricity oscillations to arise whenever $\epsGR \gtrsim 10$. This requirement severely constrains the parameter space of initial conditions that can lead to GW-assisted mergers. 

Second, fluctuations in the tidal torque felt by the binary on the timescale of its outer orbital period (which are ignored by double-averaging) can increase a binary's maximum eccentricity  \citep{Ivanov2005,Katz2012,Luo2016,Grishin2018}.  These short-timescale fluctuations (sometimes called `singly-averaged effects') can greatly enhance merger rates. Roughly speaking, one can think of them as modifying the maximum eccentricity reached by the binary from $e_\mathrm{max}$ to $\tilde{e}_\mathrm{max} = e_\mathrm{max} + \delta e$, $\delta e>0$. We take this effect into account in our calculations (see below).



\section{Calculation of the Merger fractions} 
\label{sec_fractions}


The main goal of this work is to compute the present day merger rate induced by cluster tides. Its calculation in \S \ref{sec_mergerrates} relies on knowledge of the time evolution of the \textit{merger fraction} $f_\mathrm{m}(t)$, which is found by taking a large ensemble of binaries and computing how many of them merge in a time $T_\mathrm{m}<t$. Here we outline the details of the calculation, namely, our merger time prescription (\S \ref{sec_mergertime}), the method used (\S \ref{sec_method}), and the results (\S \ref{sec_fractionresults}).


\subsection{Merger time $T_\mathrm{m}$}
\label{sec_mergertime}


An isolated binary (in the absence of cluster tides) with initial semi-major axis $a_0$ and eccentricity $e_0 \approx 1$ would merge due to GW emission in a time \citep{Peters1964}: 
\begin{align} 
T_\mathrm{m}^\mathrm{iso}(e_0) &= \frac{3c^5a_0^4}{85G^3(m_1+m_2)m_1m_2} (1-e_0^2)^{7/2}.
\label{eq:Tm_iso}
\end{align}

However, the torque from the cluster potential causes the binary's eccentricity to vary in a cyclic fashion on a secular timescale $t_\mathrm{sec}$, with $e\to 1$ under favorable circumstances. Because of the steep dependence of $T_\mathrm{m}^\mathrm{iso}$ on $1-e$, GW emission occurs in the form of discrete bursts around the sharp eccentricity maxima. Such high-$e$ episodes last for about $\Delta t_\mathrm{max}\approx t_\mathrm{sec} (1-e_\mathrm{max}^2)^{1/2}$, where $e_\mathrm{max}$ is the maximum eccentricity obtained in the DA theory (a result derived in \S 6.2 of Paper II, and routinely used in LK studies, e.g. \citealt{Miller2002}). This prolongs the time to merger (estimated using equation (\ref{eq:Tm_iso}) at peak eccentricity) by a factor $\approx t_\mathrm{sec}/\Delta t_\mathrm{max}=(1-e_\mathrm{max}^2)^{-1/2}$, see equation (\ref{eq:Tm}). 

Moreover, as $e$ passes through its peak value it also experiences short-term oscillations due to singly-averaged effects. These variations periodically take $e$ to its peak singly-averaged value $\tilde e_\mathrm{max}$, which is higher than the DA value $e_\mathrm{max}$. Again, because of the sharp dependence of GW emission on $1-e$, GW losses mainly occur when $e\approx \tilde e_\mathrm{max}$. For this reason, to approximately account for the singly-averaged effects we set the peak eccentricity determining the intensity of GW emission to $\tilde e_\mathrm{max}$ (rather than $e_\mathrm{max}$) and obtain the following estimate of the merger time: 
\ba  
T_\mathrm{m}&\approx & T_\mathrm{m}^\mathrm{iso}(\tilde e_\mathrm{max})\times (1-e_\mathrm{max}^2)^{-1/2}
\label{eq:Tm}\\  
& = &\frac{3c^5a_0^4}{85G^3(m_1+m_2)m_1m_2}\psi( e_\mathrm{max},\tilde e_\mathrm{max}) 
\label{tm}\\  
&=& 1.0 \,\mathrm{Gyr} \left( \frac{m}{1.4M_\odot} \right)^{-3} \left(\frac{a_0}{10\,\mathrm{au}}\right)^{4}  \frac{\psi(e_\mathrm{max},\tilde e_\mathrm{max})}{10^{-12}}
\nonumber \\  
&=& 0.5 \,\mathrm{Gyr} \left( \frac{m}{30M_\odot} \right)^{-3} \left(\frac{a_0}{30\,\mathrm{au}}\right)^{4} \frac{\psi(e_\mathrm{max},\tilde e_\mathrm{max})}{10^{-12}} 
\nonumber, 
\ea
where $\psi(e_\mathrm{max},\tilde e_\mathrm{max})=(1-\widetilde{e}_\mathrm{max}^2)^{7/2}(1-e_\mathrm{max}^2)^{-1/2}$. In the numerical estimates we used typical values for NS-NS and BH-BH binaries with $m_1=m_2=m$. Note that $T_\mathrm{m}$ is independent of the secular period $t_\mathrm{sec}$. 

Equation \eqref{tm} is what we use in this work for $T_\mathrm{m}$; it provides an estimate of the merger time accurate up to a factor of order unity (although see the end of \S\ref{sec_method}). A similar result for $T_\mathrm{m}$, but neglecting singly-averaged effects \footnote{We examine the impact of neglecting singly-averaged eccentriity fluctuations in Hamilton \& Rafikov (in prep.).} (i.e. with $\widetilde{e}_\mathrm{max}=e_\mathrm{max}$), has been previously used by several authors to calculate merger times of binaries driven to high eccentricity via the LK mechanism (\citealt{Thompson2011,Antonini2012,Liu2018,Grishin2018,Randall2018}).

 
\subsection{Method}
\label{sec_method}


To compute the merger fraction $f_\mathrm{m}(t)$, it is necessary that we are first able to calculate $e_\mathrm{max}$ and  $\widetilde{e}_\mathrm{max}$ for any binary.  For a given cluster potential, both $e_\mathrm{max}$ and  $\widetilde{e}_\mathrm{max}$ are functions of the eight parameters that describe the inner ($a,e,i,\omega$) and outer ($\rp,\ra$) orbits of the binary at $t=0$ and the binary component masses, e.g.
\begin{align}
e_\mathrm{max}=e_\mathrm{max}(\rp,\ra,a_0,e_0,i_0,\omega_0,m_1,m_2). \label{eq:emaxof}
\end{align} We obtain $e_\mathrm{max}$ from our secular (DA) theory including GR precession by solving equation (55) of Paper II for the value $j_\mathrm{min}=\sqrt{1-e_\mathrm{max}^2}$ at which the binary's dimensionless angular momentum $j$ reaches its minimum. Our prescription for the amplitude $\delta e$ of short-timescale fluctuations entering $\widetilde{e}_\mathrm{max}$ --- which is an approximate analytic expression similar to equation (B14) of \citet{Ivanov2005} (see also \citealt{Grishin2018}) --- is provided in Hamilton \& Rafikov (in prep.).

Then at each time $t$, for a given $a_0,m_1,m_2$ there exists a critical region in $(e_\mathrm{max},\widetilde{e}_\mathrm{max})$ space for which $T_\mathrm{m} < t$ (equation \eqref{eq:Tm}). All systems in the critical region can be considered `merged' at time $t$.  With a suitable Monte Carlo sampling of the eight parameters listed in \eqref{eq:emaxof} one can therefore compute the cumulative fraction $f_\mathrm{m}(t)$ of systems that have merged as a function of time. To carry out the Monte Carlo procedure we draw a large number\footnote{We checked that a `higher resolution' calculation which sampled $N= 10^7$ binaries gave essentially identical results.} $N= 10^6$ of binaries with initial parameters randomly chosen from appropriate distributions described as follows.

Our compact object binaries come in three flavours: NS-NS, NS-BH and BH-BH.  For the component masses $m_1,m_2$ we always use $1.4M_\odot$ (NS) and $30M_\odot$ (BH).   We use three cluster masses: $M= 10^5, 10^6, 10^7 M_\odot$. We consider two cluster potentials, the same as in Figure \ref{fig:GammaOfR}: the Plummer potential  $\Phi_\mathrm{Plum}$ to mimic cored potentials of globular clusters and the Hernquist potential $\Phi_\mathrm{Hern}$ to approximate cusped nuclear clusters. Each of them is scaled to have half mass radius  $r_\mathrm{h} = 1.31\mathrm{pc}$.

We randomly sample $\rp$ and $\ra$ (which characterize the binary's outer orbit) from a self-consistent distribution function (DF) constructed as follows. We take the isotropic self-consistent DF $g(\widetilde{\mathcal{E}}(\rp/b,\ra/b),b)$ that generates the underlying cluster potential with mass $M$ and scale radius $b$, where $\widetilde{\mathcal{E}} \equiv \mathcal{E}/(GM/b)$ and $\mathcal{E}$ is the specific energy of an orbit in that potential. Thus, $g(\widetilde{\mathcal{E}},b_\mathrm{Plum}) \propto b_\mathrm{Plum}^{-3/2}(-\widetilde{\mathcal{E}})^{7/2}$ for the Plummer potential, while for the Hernquist potential $g(\widetilde{\mathcal{E}},b_\mathrm{Hern})$ is given by equation (4.51) of \cite{Binney2008}. We then draw the orbits of our binaries from a DF  $\propto g(\widetilde{\mathcal{E}}(\rp/b^\prime,\ra/b^\prime),b^\prime)$, where the new scale radius $b^\prime$ is a parameter that we vary to account for the possibility of the massive compact object binaries being more centrally concentrated than the underlying stellar population (we leave the scale radius $b$ of the cluster potential unchanged). We choose three values of $b^\prime$ such that the corresponding central {\it over-concentration} $c \equiv \rho(0,b^\prime)/\rho(0,b)$ --- ratio of the central densities computed from the DFs $g(\widetilde{\mathcal{E}},b^\prime)$ and $g(\widetilde{\mathcal{E}},b)$ --- is equal to $1$, $10$ and $100$. Hence for $c=1$ the binaries are essentially tracer particles drawn from the underlying stellar  population, while for $c \gg 1$ they are much more centrally concentrated. In the Plummer case this requires  $b^\prime/b_\mathrm{Plum}=1,10^{-1/3}$ and $10^{-2/3}$, while for the Hernquist sphere we must take $b^\prime/b_\mathrm{Hern}=1, 10^{-1/2}$ and $10^{-1}$. Variation of $c$ helps to alleviate the observational uncertainty in the radial distribution of compact object binaries in clusters.

We assume Opik's law for the distribution of binary semi-major axes ($\md N/\md a_0 \propto a_0^{-1}$), sampling it in the range $a_0 \in (a_\mathrm{min},a_\mathrm{max})$.  Here $a_\mathrm{min}$ is the semi-major axis below which  GR precession will suppress cluster tide-driven evolution; we estimate $a_\mathrm{min}$ by solving equation \eqref{eq:epsGRnumerical} for $a$ with  $\epsilon_\mathrm{GR} = 10$ and $A^* = 1.0$. We take $a_\mathrm{max} = 50 \, \mathrm{au}, 100 \, \mathrm{au}, 100 \, \mathrm{au}$ for NS-NS, NS-BH and BH-BH binaries respectively, expecting that wider binaries would be quickly disrupted by stellar encounters.  Initial binary eccentricities are drawn from a thermal distribution (uniform in $e_0^2$) in the range $e_0 \in (0.01,0.995)$. 

We assume random orientation of the binaries, implying that the initial pericentre angles $\omega_0$ and initial cosines of inclination $\cos i_0$ are uniformly distributed in $(-\pi,\pi)$ and $(0,1)$ respectively.  However, the symmetry of the problem means that we may restrict the random sampling of  $\omega_0$ to the range $(0,\pi)$, allowing us to speed up the calculation. Moreover, only binaries with initial inclinations $i_0$ close to $90^\circ$ are able to merge within a Hubble time. This result follows from the conservation of $(1-e^2)^{1/2}\cos i$ (i.e. the $z$-component of the binary's inner orbital angular momentum, see Paper I) and the fact that very high eccentricities ($e_\mathrm{max}\to 1$) are required to enhance GW emission. Hence it is sufficient to sample $\cos i_0$ from a uniform distribution not in $(0,1)$ but $(0,\kappa)$, where we took $\kappa = 0.05, 0.08, 0.1$ for NS-NS, NS-BH and BH-BH binaries, respectively\footnote{The $\kappa$ values are calculated by putting $a_0 = a_\mathrm{min}$, $T_\mathrm{m} = 12 \, \mathrm{Gyr}$ and $1-e_\mathrm{max}^2 \sim \cos^2 i_0$ in equation \eqref{tm} and solving for $\cos i_0$ (the approximation $1-e_\mathrm{max}^2 \sim \cos^2 i_0$ is a reasonable one whenever $\Gamma > 1/5$).}.

When calculating merger fractions $f_\mathrm{m}$ we account for the aforementioned truncation of the ranges of $a_0, e_0,\omega_0,\cos i_0$. In particular we assume that the overall population of binaries has a minimum semi-major axis $0.2\,\mathrm{au}$ (whereas it is only sampled down to $a_\mathrm{min}$) while the maximum semi-major axis is still $a_\mathrm{max}$, and weight the number of merged binaries accordingly. Similarly, in reality $\cos i_0\in (0,1)$, but binaries in $(\kappa,1)$ never merge. The values of $f_\mathrm{m}(t)$ we quote always reflect the fraction of the \textit{total} population that has merged in time $t$, not just of the initial $N$ sampled binaries.

Implicit in the derivation of the merger time $T_\mathrm{m}$ is the assumption that the binary undergoes at least one secular cycle by time $t$. However, equation \eqref{tm} sometimes predicts merger times that are short compared to the secular timescale $t_\mathrm{sec}$. Since binaries must first \textit{reach} their maximum eccentricity before they can actually merge, which on average takes $\approx t_\mathrm{sec}/2$, we account for these `fast' mergers by taking the actual merger time to be $\max(T_\mathrm{m},t_\mathrm{sec}/2)$. 

 
\subsection{Merger fraction results}
\label{sec_fractionresults}


\begin{figure*}
\centering
\includegraphics[width=0.9\linewidth,trim=60 20 60 40, clip]{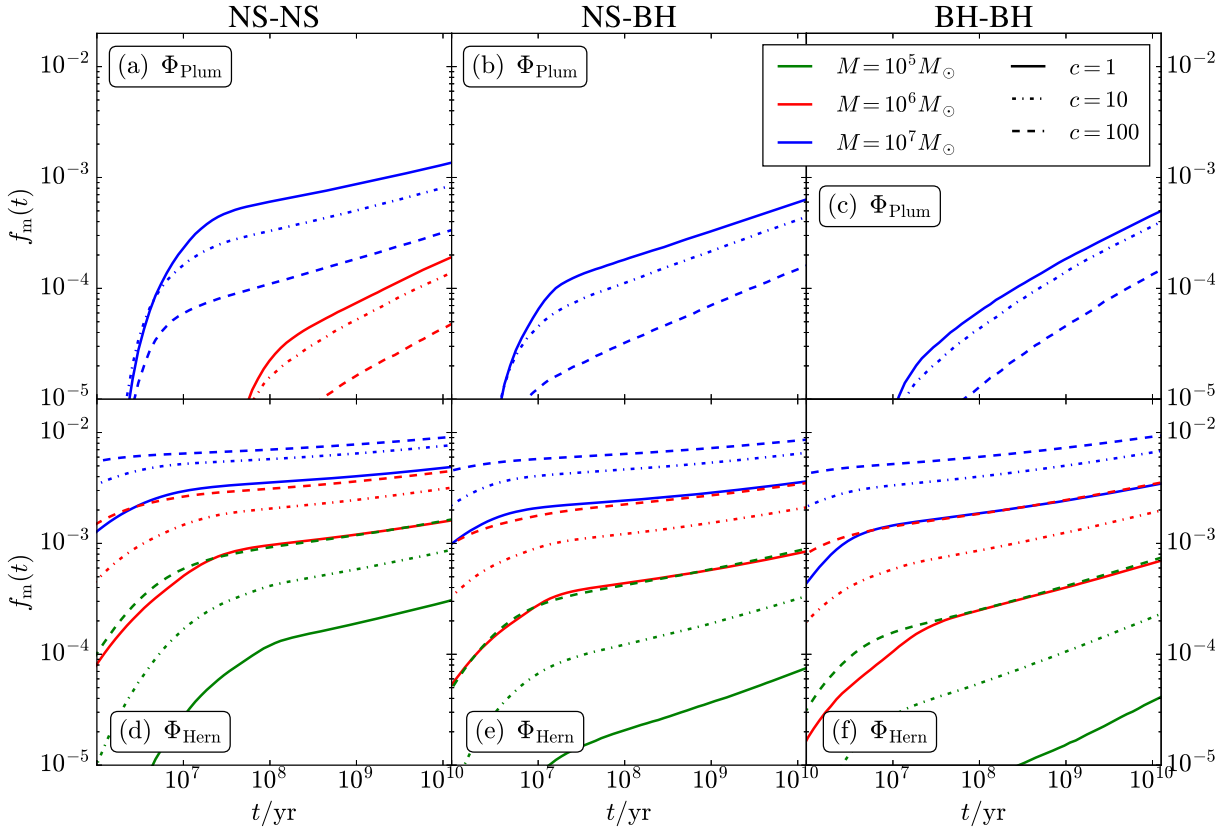}
\caption{Cumulative merger fraction $f_\mathrm{m}(t)$ over the domain $t\in (1\, \mathrm{Myr}, \, 12\, \mathrm{Gyr})$ for NS-NS, NS-BH and BH-BH binaries, each for cluster masses $M/M_\odot=10^5$, $10^6$, $10^7$ and binary central concentrations $c=1,10,100$ in the Plummer and Hernquist potentials (see legend).}
\label{fig:merger_fraction_curves}
\end{figure*}

In Figure \ref{fig:merger_fraction_curves} we plot the cumulative merger fractions $f_\mathrm{m}(t)$ for $t\in (1\,\mathrm{Myr},12\,\mathrm{Gyr})$, calculated using the method of \S \ref{sec_method}.  We consider NS-NS (left column), NS-BH (middle column), and BH-BH (right column) binaries, each for $M=10^5M_\odot$ (green), $M=10^6M_\odot$ (red) and $M=10^7M_\odot$ (blue) clusters and concentrations $c=1,10,100$ (solid, dot-dashed and dashed lines respectively), for the two potentials (\ref{eq:Plummer}).


\subsubsection{Cored (Plummer) models}
\label{eq:res_Plum}

Starting with the Plummer models (top row of Figure \ref{fig:merger_fraction_curves}), we see that $f_\mathrm{m}$ is largest for the most massive clusters ($M=10^7 M_\odot$, blue lines) because the secular evolution is fastest in such clusters and therefore large eccentricity oscillations are less easily quenched by GR precession.  For NS-NS binaries with central concentration $c=1$, the final merger fraction is $\fnow \sim 10^{-3}$ in $M=10^7 M_\odot$ clusters. The corresponding result for NS-BH and BH-BH binaries is a factor of a few smaller because of the stronger GR precession barrier for these more massive systems. In $M=10^6M_\odot$ clusters (red lines), we again find a non-negligible final NS-NS merger fraction, $f_\mathrm{m}(12 \, \mathrm{Gyr}) \sim 10^{-4}$; however, we find no NS-BH and BH-BH mergers, because for those (heavy) binaries the cluster tides are no longer strong enough to beat the GR precession. For the same reason, $f_\mathrm{m}$ is negligible in cored (Plummer) $M=10^5 M_\odot$ clusters across all binary flavours.

In all three panels, increasing the central concentration $c$ reduces the merger fraction because strongly centrally concentrated binaries in cored potentials fall into the $\Gamma < 1/5$ regime (see Figure \ref{fig:GammaOfR}) for which high eccentricity excitation is suppressed (Paper II). Mass segregation of a population of heavy binaries would act to steadily increase $c(t)$ over the age of the cluster.  In cored clusters this would lead to a lower merger fraction at late times compared to an unsegregated population.


\subsubsection{Cusped (Hernquist) models}
\label{eq:res_Hern}

Cusped clusters represented by a Hernquist potential (bottom row of Figure \ref{fig:merger_fraction_curves}) exhibit substantially higher $f_\mathrm{m}$ values than in the Plummer case. Indeed, even $10^5 M_\odot$ clusters (green curves) --- which produced zero mergers in the Plummer potential --- now have $\fnow$ of at least a few $\times \, 10^{-5}$ and often as large as $\sim 10^{-3}$, depending on $c$ and the binary type. Moreover, increasing $c$ in these potentials \textit{increases} $f_\mathrm{m}$, which is the opposite trend to the Plummer case. As a result, mass segregation in cusped clusters would tend to additionally increase $f_\mathrm{m}$ at late times.

Both effects are due to the ubiquity of the $\Gamma>1/5$ regime (promoting high $e$ excitation) in the Hernquist potential, even near the cluster centre (Figure \ref{fig:GammaOfR}) --- unlike in the Plummer case, there is little disadvantage to binaries being centrally concentrated. Moreover, secular evolution is fast near the centre of the Hernquist sphere ($t_\mathrm{sec}\propto A^{-1}$ and the `tidal strength' $A$ diverges, see Figure \ref{fig:GammaOfR}), and short-timescale fluctuations there are strong.  As a result, increasing $c$ drives more binaries to merge within a Hubble time.  Many binaries that orbit near the centres of cuspy clusters have $t_\mathrm{sec}< 10^6 \mathrm{yr}$ --- hence, several curves show nonzero $f_\mathrm{m}(10^6 \, \mathrm{yr})$.

Also, $f_\mathrm{m}$ shows a weaker dependence on cluster mass $M$ than in the Plummer case.  This is because of the large $A$ values in the Hernquist case (see Fig. \ref{fig:GammaOfR}b), which act to suppress the effect of GR precession: equation \eqref{eq:epsGRformula} then yields $\epsGR\to 0$, a limit in which $e_\mathrm{max}$ is independent of $M$ (Paper II).


\section{Merger rates}
\label{sec_mergerrates}


Our results on merger fractions $f_\mathrm{m}(t)$ allow us to calculate the \textit{specific merger rate} $\mathcal{R}$, which is the rate of compact object binary mergers of a given flavour per unit volume in the local universe, given the birth history of binaries of that type. The latter is described by the formation rate of such binaries per unit cluster mass $W(t)$, such that in the interval $(t,t+\delta t)$, $W(t)\delta t$ systems are produced per unit cluster mass. The cumulative number of mergers from that binary type \textit{per unit cluster mass} after time $t$ is then  
\begin{align} 
\mathcal{C}(t) \equiv \int_0^t \md t' W(t') f_\mathrm{m}(t-t'),
\end{align}  
and the corresponding contribution to the specific merger {\it rate} at time $t$ is $\mathcal{R}=\rho_\mathrm{cl}\md \mathcal{C}(t)/\md t$, where $\rho_\mathrm{cl}$ is the cluster mass density in the local universe.

We consider two simple histories of compact object binary formation. The first takes the form of a burst, so that at $t=0$ each cluster instantaneously forms a population of binaries. If $X_\mathrm{born}$ compact object binaries are born per unit cluster mass, then $W(t)=X_\mathrm{born}\delta(t)$ so that $\mathcal{C}(t) = X_\mathrm{born}f_\mathrm{m}(t)$ and
\ba    
\mathcal{R}(t)=X_\mathrm{born}\rho_\mathrm{cl}\frac{\md f_\mathrm{m}(t)}{\md t}.
\label{eq:Rburst}
\ea  

The second model assumes a constant compact object binary formation rate $W(t)=Y_\mathrm{form}$ per unit cluster mass.  Then the cumulative merger number from that cluster is $\mathcal{C}(t) = Y_\mathrm{form} \int_0^t \md t' f_\mathrm{m}(t-t') = Y_\mathrm{form} \int_0^t \md x \,f_\mathrm{m}(x)$, resulting in the specific merger rate
\ba    
\mathcal{R}(t)=Y_\mathrm{form}\rho_\mathrm{cl}f_\mathrm{m}(t).
\label{eq:Rflat}
\ea  

The results obtained for these two binary formation histories give an idea of the outcomes of more sophisticated models.


\subsection{Merger rates from globular clusters}
\label{sec:gc_rate}


Globular clusters have cored profiles, so we use $f_\mathrm{m}$ results for Plummer spheres (\S \ref{eq:res_Plum}) to represent them. Since globulars have a range of masses and $f_\mathrm{m}$ is a function of $M$, appropriate averaging of the rates (\ref{eq:Rburst})-(\ref{eq:Rflat}) over the cluster mass spectrum is needed. Following \citet{Rodriguez2015} we use a log-normal mass function for the number density of globulars \citep{Harris2014}:
\ba
    \frac{\md n_\mathrm{gc}}{\md \lg(M/M_\odot)} && = \frac{n^\mathrm{tot}_\mathrm{gc}}{\sqrt{2\pi}\sigma_M}
    \nonumber\\
    && \times \exp\left[-\frac{(\lg(M/M_\odot)-\mu)^2}{2\sigma_M^2} \right], \label{eq:Mdist}
\ea
where $n^\mathrm{tot}_\mathrm{gc}$ is the total number density of globular clusters in the local universe integrated over $M$, and $\sigma_M = 0.52$, $\mu = 5.54$. The number density $n_\mathrm{gc}^\mathrm{tot}$ is an uncertain quantity \citep{PortegiesZwart2000,Rodriguez2015,Rodriguez2016a}. In this work, guided by existing estimates, we adopt $n_\mathrm{gc}^\mathrm{tot} = 3 \,\mathrm{Mpc}^{-3}$. 

For simplicity, we do the averaging in an approximate fashion by splitting the cluster population into 3 mass bins $M_{i}^\mathrm{min}<M<M_{i}^\mathrm{max}$, $i=1,2,3$, where $M_{i}^\mathrm{min}=5\times 10^{3+i}M_\odot$ and $M_{i}^\mathrm{max}=5\times 10^{4+i}M_\odot$. The mass density in clusters in each mass bin is then
$\rho_{\mathrm{gc},i}=\int_{M_{i}^\mathrm{min}}^{M_{i}^\mathrm{max}}M\md n_\mathrm{gc}=(3.9,14.1,3.3)\times 10^5 (n^\mathrm{tot}_\mathrm{gc}/3~\mathrm{Mpc}^{-3})M_\odot$ Mpc$^{-3}$. We assign to each bin the value of $f_\mathrm{m}$ computed for Plummer models with $M=M_i=10^{4+i}M_\odot$ (within the $i$-th bin). Then averaging of the merger rate over the distribution of $M$ amounts to replacing $\rho_\mathrm{cl}f_\mathrm{m}$ with 
\ba   
F_\mathrm{m}(t)=\sum\limits_{i=1}^3\rho_{\mathrm{gc},i}f_\mathrm{m}(t;M_i).
\label{eq:Fm}
\ea  

We now compute the present day rate $\mathcal{R}$ for the two aforementioned binary birth histories.



\subsubsection{Merger rates from globular clusters: a single burst of compact object binary formation}
\label{sec:gc_rate_single_burst}

Globular clusters experience a large starburst at their formation. Compact objects get produced in supernova explosions shortly thereafter. If they remain bound and assemble into binaries on a timescale short compared to the Hubble time, then the single burst approximation (\ref{eq:Rburst}) should characterize the current merger rate $\mathcal{R}$ reasonably well.  

Motivated by the calculations of \citet{Lockmann2010}, in this work we adopt $X_\mathrm{born} = 10^{-3}M_\odot^{-1}$ for the specific birth rate of all compact binary species, similar to the value obtained in \citet{Rodriguez2016a}. We calculate the total merger rate using equation (\ref{eq:Rburst}), averaging it over cluster mass via equation (\ref{eq:Fm}): 
\begin{align}
\mathcal{R} &= X_\mathrm{born} \frac{\md F_\mathrm{m}}{\md t}= 3\times 10^{-3} \, \mathrm{Gpc}^{-3}\mathrm{yr}^{-1}  \\
& \times\frac{X_\mathrm{born}}{10^{-3}M_\odot^{-1}}~ 
\frac{\md F_\mathrm{m}/\md  t \vert_{12\,\mathrm{Gyr}}}{3 ~M_\odot~\mathrm{Mpc}^{-3}\mathrm{Gyr}^{-1}}, \nn
\label{eqn_Rtot1}
\end{align} 
where in the numerical estimate we assumed that the formation burst happened 12 Gyr ago, and took a value of $\md F_\mathrm{m} /\md t$ characteristic of Plummer models (\S \ref{eq:res_Plum}).


\subsubsection{Merger rates from globular clusters: a constant rate of compact object binary formation}
\label{sec:gc_rate_constant_formation}

An alternative birth history is the one in which the assembly of compact objects into binaries in globulars occured at a steady (slow) rate $Y_\mathrm{form}$ over the last $12\,\mathrm{Gyr}$. Here we adopt $Y_\mathrm{form}=10^{-4}M_\odot^{-1}\mathrm{Gyr}^{-1}$ so that upon integration over a Hubble time we reproduce roughly the specific compact binary occurrence rate $X_\mathrm{born}$ assumed in \S \ref{sec:gc_rate_single_burst} (i.e. $Y_\mathrm{form}\times 10$ Gyr $=X_\mathrm{born}$). Then from equation (\ref{eq:Rflat}) the merger rate is   
\begin{align}
\mathcal{R} &= Y_\mathrm{form} F_\mathrm{m}= 0.3 \, \mathrm{Gpc}^{-3}\mathrm{yr}^{-1} \\
& \times \frac{Y_\mathrm{form}}{10^{-4}M_\odot^{-1}\mathrm{Gyr}^{-1}} ~\frac{F_\mathrm{m}(12 \, \mathrm{Gyr})}{3\times 10^3 ~M_\odot~\mathrm{Mpc}^{-3}},\nn
\label{eqn_Rtot2}
\end{align}
and again we took $F_\mathrm{m}(12\,\mathrm{Gyr})$ values characteristic of Plummer models (\S \ref{eq:res_Plum}).


\subsection{Merger rates from nuclear clusters}
\label{sec:nc_rate}


In the case of nuclear clusters we expect compact object binaries to be created at a relatively steady rate due to continuous star formation over long times (\citealt{Figer2004}; dynamical assembly due to 3-body processes is not as important here, although see \citealt{Muno2005}). Thus, the constant formation rate assumption is more appropriate for nuclear clusters, and we again assume $Y_\mathrm{form}=10^{-4}M_\odot^{-1}\mathrm{Gyr}^{-1}$ for these systems. 

For simplicity, we take all nuclear clusters to have mass $M_\mathrm{nc}=10^7 M_\odot$ and assume $n_\mathrm{nc} = 0.02~ \mathrm{Mpc}^{-3}$ for their number density \citep{Petrovich2017,Hamers2018_VRR}. Then $\rho_\mathrm{cl} f_\mathrm{m}=M_\mathrm{nc}n_\mathrm{nc}f_\mathrm{m}(M_\mathrm{nc})$ and the merger rate becomes (equation (\ref{eq:Rflat}))
\begin{align}
\mathcal{R} &= Y_\mathrm{form} M_\mathrm{nc}n_\mathrm{nc}f_\mathrm{m}(M_\mathrm{nc})\\
& = 0.2 \, \mathrm{Gpc}^{-3}\mathrm{yr}^{-1}  ~\frac{Y_\mathrm{form}}{10^{-4}M_\odot^{-1}\mathrm{Gyr}^{-1}}\nn\\ 
&\times \frac{n_\mathrm{nc}}{0.02 \mathrm{Mpc}^{-3}}~
\frac{f_\mathrm{m}(12\, \mathrm{Gyr}; M_\mathrm{nc})}{10^{-2}}
\nn, 
\label{eqn_Rtotnuc}
\end{align} 
where for $f_\mathrm{m}(12\, \mathrm{Gyr};M_\mathrm{nc})$ we adopted a value characteristic of cusped (Hernquist) models --- see \S \ref{eq:res_Hern}.  Cored nuclear clusters have $f_\mathrm{m}(12\, \mathrm{Gyr}; M_\mathrm{nc})$ an order of magnitude lower, see \S \ref{eq:res_Plum}.

\begin{figure*}
\centering
\includegraphics[width=0.92\linewidth,trim=65 10 65 0, clip]{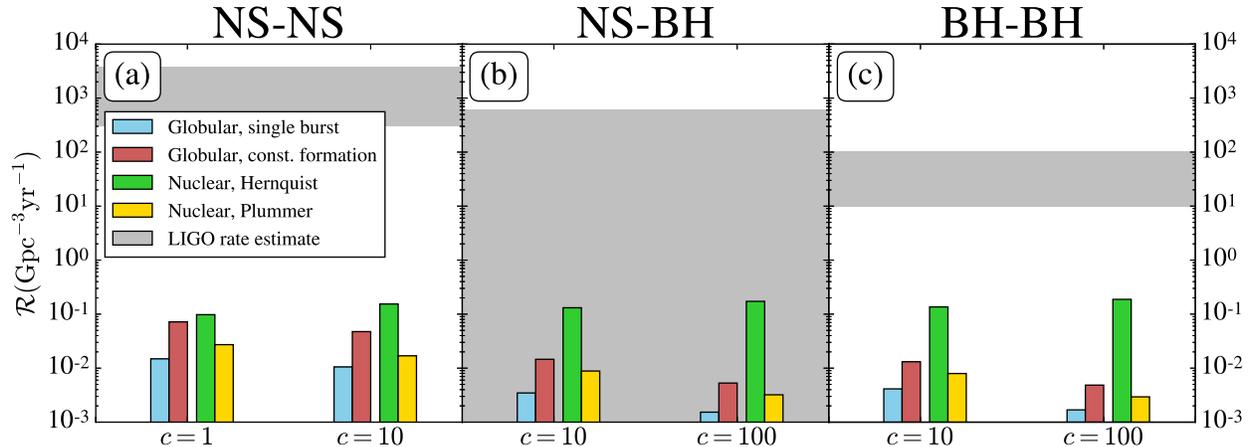}
\caption{Merger rates $\mathcal{R}$ of compact object binaries driven by the tidal fields of (spherical) globular and nuclear clusters. For each binary type we consider two values of the central concentration $c$. For globular clusters, modeled as cored (Plummer) systems, we look at two binary birth histories: single burst (blue) and constant formation rate (red). For nuclear clusters we calculate rates assuming either cored (Plummer, yellow) or cusped (Hernquist, green) profiles, considering only the constant binary formation history. Grey regions show the LIGO rate estimates. See text for details.}
\label{fig:mergerrates}
\end{figure*}


\section{Discussion}
\label{sec_discussion}


In Figure \ref{fig:mergerrates} we show present day compact binary merger rates due to cluster tides in globular and nuclear clusters. Rates for globulars use the results we obtained for Plummer models (for two birth histories, \S\S \ref{sec:gc_rate_single_burst}-\ref{sec:gc_rate_constant_formation}), while for nuclear clusters we consider both Hernquist and Plummer models and a flat binary formation history (\S \ref{sec:nc_rate}). For NS-NS binaries we consider only moderate concentrations $c=1,10$, while for (significantly heavier) NS-BH and BH-BH binaries we assumed a higher degree of central segregation, $c=10,100$. The grey regions in Figure \ref{fig:mergerrates} show the LIGO rate estimates \citep{LIGO2018}: $110-3840 \, \mathrm{Gpc}^{-3} \, \mathrm{yr}^{-1}$ and $9.7-101 \, \mathrm{Gpc}^{-3} \, \mathrm{yr}^{-1}$ for NS-NS and BH-BH mergers in the local universe respectively, while the upper limit on the NS-BH merger rate is $610 \, \mathrm{Gpc}^{-3} \, \mathrm{yr}^{-1}$. 

Focusing first on globular clusters, one can see that their merger rates fall short of providing a substantial contribution to the observed rates. We find $\mathcal{R}\sim 0.01 - 0.07 \, \mathrm{Gpc}^{-3}\mathrm{yr}^{-1}$ for NS-NS binaries and $\lesssim 0.02 \, \mathrm{Gpc}^{-3}\mathrm{yr}^{-1}$ for each of NS-BH and BH-BH binaries in globulars (Figure \ref{fig:mergerrates}).  The primary reason for fewer NS-BH and BH-BH mergers compared to NS-NS mergers is that the heavier binaries (i) suffer from stronger GR precession which cannot be overcome in a cored potential even at the cluster center, (ii) have higher central concentrations which brings them in to the $\Gamma < 1/5$ regime, where high eccentricity excitation is suppressed (higher $c$ always leads to lower $\mathcal{R}$ in globulars). Also, a constant binary formation rate results in higher $\mathcal{R}$ because many binaries merge soon after their birth: $f_\mathrm{m}(t)$ curves rise substantially faster during the first $10^7-10^8$ yr, see Figure \ref{fig:merger_fraction_curves}.

As for nuclear star clusters, if we assume a cusped density profile (Hernquist model) then $\mathcal{R}\sim 0.1-0.2 \, \mathrm{Gpc}^{-3}\mathrm{yr}^{-1}$ for NS-NS, NS-BH and BH-BH binaries.  The NS-BH and BH-BH binaries merge \textit{slightly} more often than NS-NS binaries because near the centre of cusped clusters the $\Gamma < 1/5$ regime is rare, and the tidal field is strong which helps to overcome GR precession. As a consequence, higher central concentration is advantageous (although not dramatically). However, in cored nuclear clusters the situation is more similar to that in globulars and $\mathcal{R}$ drops appreciably with increasing $c$.

Overall, we see that NS-BH and BH-BH merger rates are very similar, assuming they are formed in equal numbers. Cusped nuclear clusters dominate the cluster tide-driven merger rate compared to globulars for all binary species. Whereas cluster tides acting alone are unlikely to produce many NS-NS mergers, they can still contribute at the level of several per cent to the observed NS-BH and BH-BH merger rates, given our assumptions.


 
\subsection{Comparison with existing studies}
\label{sec:litcomparison}


There are a number of existing estimates of compact object binary merger rates in globular and nuclear clusters \citep{Antonini2014,Stephan2016, Antonini2016b,Fragione2019b}. The studies which bear closest resemblance to our work consider binaries orbiting SMBHs at the centres of spherical nuclear clusters and undergoing LK-driven evolution \citep{Antonini2012,Prodan2015,Hoang2018}. \citet{Petrovich2017} explored a similar setup (binary orbiting a SMBH) but also included the effect of a \textit{non}-spherical cluster potential on the orientation of the binary's \textit{outer} orbit. As a result of nodal precession of the outer orbit, the inclination of the inner binary (with respect to its outer orbit) was able to reach high values, triggering LK oscillations and greatly enhancing merger rates.

However, none of these studies accounted for the direct tidal torque on the inner orbit due to the cluster potential as we do here. Additionally, in these studies the distribution of binary outer orbits is typically truncated at radii of $\lesssim 0.1 \mathrm{pc}$ from the cluster centre. We do not rely on the presence of a central black hole and still find mergers (out to much larger radii) by including a cluster potential. 

In nuclear clusters our BH-BH merger rate $\mathcal{R} \sim 0.1 - 0.2 \, \mathrm{Gpc}^{-3}\mathrm{yr}^{-1}$ is comparable to (but typically slightly smaller than) those of others, e.g. \citet{Antonini2016a} ($\mathcal{R} \sim 1 \, \mathrm{Gpc}^{-3}\mathrm{yr}^{-1}$ from nuclear clusters without a SMBH), \citet{Petrovich2017} ($\mathcal{R} \sim 0.6 - 15 \, \mathrm{Gpc}^{-3}\mathrm{yr}^{-1}$ from non-spherical nuclear clusters with a SMBH, but they use higher $Y_\mathrm{form}$).  In globulars our BH-BH rate $\mathcal{R} \lesssim 0.02 \, \mathrm{Gpc}^{-3}\mathrm{yr}^{-1}$ is significantly smaller than those of e.g. \citet{Rodriguez2016a} ($\mathcal{R} \sim 2-20 \, \mathrm{Gpc}^{-3}\mathrm{yr}^{-1}$ from hardening of dynamically formed binaries), see \S \ref{sec:gc_rate_constant_formation}.

For NS-BH and NS-NS binaries in (cusped) nuclear clusters our rates, $\mathcal{R} \sim 0.1 - 0.2 \, \mathrm{Gpc}^{-3}\mathrm{yr}^{-1}$, are comparable to or greater than those of \citet{Petrovich2017} ($\mathcal{R} \sim 0.02 - 0.4 \, \mathrm{Gpc}^{-3}\mathrm{yr}^{-1}$ and $\mathcal{R} \lesssim 0.02 \,\mathrm{Gpc}^{-3}\mathrm{yr}^{-1}$ respectively).  Our results are also comparable to those of \citet{Hamers2018_VRR} who found a combined merger rate for all compact object binary flavours in nuclear clusters with SMBHs of $\mathcal{R} \sim 0.02 - 0.4 \, \mathrm{Gpc}^{-3}\mathrm{yr}^{-1}$.

Like most other dynamical merger channels, the rates produced by our mechanism fall short of those observed by LIGO by at least one order of magnitude.

 
\subsection{Further refinements}
\label{sec:limitations}


Apart from some technical simplifications used in this study (e.g. our approximation of $T_\mathrm{m}$ using equation \eqref{tm}, simple analytical estimate for $\delta e$, etc.), we have also deliberately omitted certain physical ingredients to focus on mergers arising due to secular effects alone. 

Perhaps most crucially, we ignored the impact of flyby encounters on the binary's inner orbital elements \citep{Heggie1996,Hamers2018_flybys}. This is an important effect that can influence our results in non-trivial ways. Recently, \cite{Samsing2019} found that numerous distant flybys can systematically increase the number of binary mergers in stellar clusters (although they did not account for secular tide-driven evolution). \citet{Heisler1986}, in their study of Oort comet dynamics, found that stellar flybys contribute a significant portion of the torque at high eccentricity --- in fact, the Oort comets exhibit a coupled behavior in which their
orbital elements roughly follow a smooth, secular (Galactic tide-driven) trajectory on average, while simultaneously exhibiting a random walk in phase-space because of stochastic flyby encounters. We expect a similar behaviour to hold in our case, and will explore it in future work.

We also neglected time-dependence of the cluster properties, e.g. due to core collapse or disk shocking, and ignored the relaxation of the binary's outer orbit e.g. due to vector resonant relaxation \citep{VanLandingham2016,Hamers2018_VRR} or dynamical friction. In particular, mass segregation of heavy binaries would boost the central concentration $c$, which can increase or decrease merger fractions depending on the cluster potential and the level of concentration (\S\ref{sec_fractions}). However, we note that our merger rates are often only mildly  affected by variation of $c$ (Figure \ref{fig:mergerrates}).

To focus on the tidal effect of the smooth cluster mass distribution alone, in this work we ignored the possibility of a central SMBH which could reside in nuclear clusters. Similarly, we assumed each cluster to be perfectly spherically symmetric, omitting the effects of possible oblateness on the outer orbit \citep{Petrovich2017}. 

Our future work will address many of these issues. In Hamilton \& Rafikov (in prep.) we will explore the sensitivity of our results to variation of the underlying assumptions, and study the impact of the presence of a central SMBH on the merger rates in nuclear clusters. 

 
\subsection{Summary}
\label{sec:sum}


We explored a new channel for producing compact object mergers in dense stellar clusters which relies on the secular evolution of binaries driven by the cluster's tidal gravitational field (a field which is unavoidably present in any merger model involving clusters). We computed merger rates due to this mechanism by focusing on conditions in which the binary can be driven to such high eccentricity that GW emission becomes important, while fully accounting for the detrimental effect of GR precession. We showed that stellar systems with cored potentials (e.g. globular clusters) do not produce many mergers, owing to the inefficiency of high-eccentricity excitation in the cluster cores. Cusped nuclear clusters (even in the absence of a central SMBH) are significantly more effective and lead to observationally interesting merger rates. Our merger rates come closest to meeting current LIGO estimates for BH-BH binaries but still fall short by more than an order of magnitude.  On the other hand, we note that all current rate estimates --- including ours --- have (systematic) error bars of at least an order of magnitude. Our future work will refine these calculations in many ways.

\acknowledgements 

CH is funded by a Science and Technology Facilities Council (STFC) studentship.




\bibliographystyle{apj}
\bibliography{Bibliography} 


\end{document}